\documentclass[intlimits,twoside,a4paper]{article}

\usepackage{amsmath,amssymb}
\usepackage{graphicx}
\usepackage{wrapfig}

\usepackage[T2A]{fontenc}
\usepackage[cp1251]{inputenc}

\usepackage{color}

\usepackage{cmpj2}

\issue{2012}{15}{1}{13706}

\doinumber{10.5488/CMP.15.13706}

\title[The high-temperature expansion of the classical Ising model
with $S_z^2$ term]{The high-temperature expansion of the classical Ising model
with $S_z^2$ term}

\author[M.T. Thomaz, O.~Rojas]{M.T. Thomaz\refaddr{ad1}\footnote{E-mail: mtt@if.uff.br}\,, O.~Rojas\refaddr{ad2}}

\addresses{\addr{ad1}Instituto de F\'{\i}sica, Universidade Federal Fluminense, Av. Gal. Milton Tavares de Souza s/n$^{\textit o}$,
CEP 24210--346, Niter\'oi-RJ, Brazil
\addr{ad2}Departamento de Ci\^encias Exatas, Universidade Federal de Lavras, Caixa Postal 3037, CEP 37200--000, Lavras-MG,  Brazil
}

\date{Received December 28, 2011, in final form February 13, 2012}

\authorcopyright{M.T. Thomaz, O.~Rojas, 2012}

\begin{document}

\maketitle

\begin{abstract}

We derive the high-temperature expansion of the Helmholtz free
energy  up to the order $\beta^{17}$ of the
one-dimensional spin-$S$ Ising model, with single-ion
anisotropy term, in the \mbox{presence}  of a longitudinal
magnetic field. We show  that the
values of some thermodynamical functions for the
ferromagnetic models, in the presence of a
weak magnetic field, are not small corrections to
their values with $h=0$.  This model with $S=3$ was applied by
Kishine  et al. [J.-i. Kishine et al.,
Phys. Rev.~B, 2006, \textbf{74}, 224419] to analyze
experimental data  of the single-chain magnet
${\rm [Mn (saltmen)]_2 [Ni(pac)_2 (py)_2] (PF_6)_2}$ for $T<40$~K.
We show that for $T<35$~K the  thermodynamic  functions  of
the large-spin limit model are poor approximations to their analogous
spin-$3$ functions.

\pacs 75.10.Jm, 05.30.-d, 75.50.Xx

\end{abstract}

\section{Introduction} \label{S1}

The one-dimensional spin-$1/2$ Ising model with
first-neighbor interaction, in the presence of a longitudinal
magnetic field, was exactly solved in 1925~\cite{ising}.
The Helmholtz free energy (HFE) of this model with $S=1/2$ has a simple
mathematical expression~\cite{baxter}.
The exact expression of the HFE for the $S=1$ ferromagnetic
model  in the presence of a longitudinal magnetic field
was derived in 1976 by Krinsky and  Furman~\cite{krinsky}
using the matrix density approach. (This work seems to have been
neglected by the subsequent literature, though). More recently,
the HFE of the $S=1$~\cite{mancini,mancini2008} and $S= 3/2$~\cite{avella}
of the Ising model in the presence of an external magnetic field
have been written as a set of coupled equations and solved
numerically.

In 2007  Rojas  et al.~\cite{phys07} published the high-temperature expansion
of the HFE of the Ising model for arbitrary value $S$ of the spin,
in the absence of a magnetic field, up to order $\beta^{40}$. The nice
feature of such expansion is that the value $S$ of the spin is an
arbitrary parameter ($S = 1/2,\, 1, \,3/2, \ldots$). The absence of an external
magnetic field in the model discussed in~\cite{phys07}
yields no distinction in the
behavior of some thermodynamical functions
between the ferromagnetic and the anti-ferromagnetic (AF)
case, e.g. the specific heat per site.

An active area of molecular chemistry is that of
designing new magnetic materials that present a
one-dimensional nanomagnetic behaviour with a strong
anisotropy axis~\cite{z-h}.
Some molecular single-chain magnets (SCMs) exhibit a strong
uniaxial (Ising) anisotropy. Kishine   et al.~\cite{kishine}
applied  the Blume-Capel model~\cite{blume,capel} (the one-dimensional
Ising model with the single-ion anisotropy term) with $S=3$ in the
presence of a longitudinal magnetic field to analyze the low-energy
dynamics response of the SCM
${\rm [Mn (saltmen)]_2 [Ni(pac)_2 (py)_2] (PF_6)_2}$ for
temperature $T\lesssim 40$~K. They concluded that the
experimental data support the view that this SCM can be described, in this
window of temperature, by the spin-$3$ of this model.

We apply the method of cumulants described in~\cite{chain_m}
to calculate the high-temperature expansion of the HFE of any one-dimensional
Hamiltonian which is invariant under space translation  and satisfies
the periodic space condition. In this approach, in order to obtain  the exact
coefficient that multiplies the term of order $\beta^n$ in the expansion
we have to calculate a set of functions named $H_{1,m}^{(n)}$,
$m = 1, 2, \ldots, n$.  The interested reader will find a survey
of the method in reference~\cite{winder}.
In section~\ref{S3} we apply the results of reference~\cite{chain_m}
to  calculate  the  high-temperature expansion  of the HFE of the
normalized one-dimensional spin-$S$ Ising model with
single-ion anisotropy term in the presence of an external
longitudinal magnetic field up to order $\beta^{17}$.
The nice feature about this expansion is that
it is faster to apply than the transfer-matrix
method~\cite{baxter}, although the
latter one provides exact curves in the whole interval of
temperatures.  In order to show the importance
of calculating the high-temperature expansion
of the spin-$S$ Ising model in the presence of an external
magnetic field, in section~\ref{S4} we compare
the behavior of certain thermodynamical functions of the
ferromagnetic and of the AF Blume-Capel models
in the presence of a weak longitudinal magnetic field.
This comparison is made for various values of the spin, including
the large-spin limit ($S\rightarrow \infty$).  Section~\ref{S5}
presents the thermodynamics of the SCM (for spin-$3$ and the
large-spin limit model)  in the interval
$11~\text{K} \lesssim T \lesssim 40~\text{K}$.
Finally, in section~\ref{S6} we present our conclusions.


\section{Thermodynamics of the spin-$S$ Ising model with
a single-ion \\ aniso\-tropy term} \label{S3}

The Hamiltonian of the spin-$S$ Ising model with a single-ion
anisotropy term in the presence of a longitudinal magnetic field
is~\cite{winder}

\begin{eqnarray} \label{3.1}
{\mathbf H}_S^{\prime} = \overset{N}{\underset{i=1} \sum} \;\;
\left[ J^{\prime} S_i^z S_{i+1}^z - h^{\prime} S_i^z
   + D^{\prime} \left(S_i^z\right)^2 \right] ,
\end{eqnarray}

\noindent where $S_i^z$ is the $z$ component  of the
spin $\vec{S}$ operator  with norm: $||\vec{S}||^2 =  S(S+1)$,
$S = \frac{1}{2}, 1, \frac{3}{2}, 2, \ldots$, at the $i$-th site
of the chain; $J^{\prime}$ is the exchange strength and it can have
negative value (ferromagnetic model) or positive value (AF
model). In references~\cite{bjp01} and~\cite{winder} we studied
the exact thermodynamics of this model for $S= \frac{1}{2}$ and
$1$ (with $h=0$), respectively.

If the large-spin limit ($S \rightarrow \infty$) is
applied directly to the Hamiltonian~(\ref{3.1}), all
the  thermodynamic functions of the large-spin limit model will diverge.
In order to keep the functions finite in this limit, we study the
normalized version of the Hamiltonian~(\ref{3.1}),
that is,
\begin{eqnarray}  \label{3.2}
{\mathbf H}_s = \overset{N}{\underset{i=1} \sum} \,\,
\left[J s_i^z s_{i+1}^z  -  h s_i^z  + D \left(s_i^z\right)^2\right] ,
\hspace{0.5cm} s = \frac{1}{2}, 1, \frac{3}{3}, 2, \ldots ,
\end{eqnarray}
where $s_i^z$ is the $z$ component of the spin
operator $\vec{s}$ that has norm 1. The $\vec{s}$ operator
is defined as
\begin{eqnarray}  \label{3.3}
\vec{s}  \equiv \frac{\vec{S}}{\sqrt{S (S+1)}} ,
\hspace{0.5cm}  S = \frac{1}{2}, 1, \frac{3}{2}, 2, \ldots \,.
\end{eqnarray}
Making $S \rightarrow \infty$ in the Hamiltonian
(\ref{3.2}) we obtain  its large-spin limit.

The Hamiltonians (\ref{3.1}) and (\ref{3.2}) are identical
when
\begin{eqnarray}  \label{3.4}
J = S(S+1) J^{\prime}, \qquad  h = \sqrt{S (S+1)} h^{\prime}
\qquad \text{and} \qquad
D = S(S+1) D^{\prime}
\end{eqnarray}
with $S = \frac{1}{2}, 1, \frac{3}{2}, 2, \ldots$\,.

Let ${\mathcal Z}_S^{\prime} (J^{\prime}\!, h^{\prime}\!, D^{\prime}\!; \beta)$
be the partition function derived from Hamiltonian~(\ref{3.1}),
${\mathcal Z}_S^{\prime} (J^{\prime}\!, h^{\prime}\!, D^{\prime}\!; \beta)=\mathrm{Tr} \left(\re^{-\beta {\mathbf H}_S^{\prime}}\right) $.
Its HFE, in the thermodynamic limit, is called ${\mathcal W}_S^{\prime}$,
where
\begin{eqnarray}  \label{3.6}
{\mathcal W}_S^{\prime} (J^{\prime}, h^{\prime}, D^{\prime}; \beta)
= -  {\underset {N \rightarrow \infty} \lim} \; \frac{1}{N} \; \frac{1}{\beta}
\;\; \ln \left[{\mathcal Z}_S^{\prime} (J^{\prime}, h^{\prime}, D^{\prime}; \beta) \right] .
\end{eqnarray}

The analogous functions for the normalized Hamiltonian (\ref{3.2}) are
${\mathcal Z}_s (J, h, D; \beta)
= \mathrm{Tr} \left( e^{-\beta {\mathbf H}_s} \right)$,
and
\begin{eqnarray}  \label{3.7b}
{\mathcal W}_s (J, h, D; \beta)
= -  {\underset {N \rightarrow \infty} \lim} \; \frac{1}{N} \; \frac{1}{\beta}
\;\; \ln \left[{\mathcal Z}_s (J, h, D; \beta) \right] .
\end{eqnarray}

For arbitrary spin value $S$, the relation between the  HFE's (\ref{3.6})
and (\ref{3.7b}) is
\begin{eqnarray}   \label{3.8}
{\mathcal W}_S^{\prime} (J^{\prime}, h^{\prime}, D^{\prime}; \beta) =
{\mathcal W}_s \big( S(S+1) J^{\prime}, \sqrt{S(S+1)} h^{\prime}, S(S+1) D^{\prime}; \beta\big) .
\end{eqnarray}

The Hamiltonian (\ref{3.1}) and its normalized version (\ref{3.2})
belong to a subclass of chain Hamiltonians that satisfy
periodic conditions  and  that can be decomposed as
${\mathbf H}_{i, i+1} = {\mathbf P}_{i, i+1} + {\mathbf Q}_i$ ,
in which the operator ${\mathbf P}_{i, i+1}$ depends on two sites (the $i$-th and the
$(i+1)$-th sites), the operator ${\mathbf Q}_i$ depends only on the
$i$-th site; moreover,
\begin{eqnarray}  \label{2.8}
 \mathrm{tr}_i \left({\mathbf P}_{i, i+1}\right) = 0 .
\end{eqnarray}

It is simple to show  that the functions $H_{1,m}^{(n)}$
calculated in the method presented in references~\cite{chain_m,winder}
for the subclass of Hamiltonians satisfying the
condition~(\ref{2.8}) can be written as
\begin{eqnarray}   \label{2.10}
H_{1,m}^{(n)} = H_{1,1}^{(1)} \times H_{1, m-1}^{(n-1)}  +
\overset{n} {\underset{n_i \not=  0, \;\; i = 1, \,\ldots, \, m}
{\underset{n_1+ n_2 +n_3 + \ldots + n_m = n} {\underset{n_2, \,n_3, \, \ldots, \, n_m = 1} {\underset{n_1 =2} \sum}}  } }
\;\; \left\langle \frac{{\mathbf H}_{12}^{n_1}}{ n_1 !} \; \frac{{\mathbf H}_{23}^{n_2}}{ n_2 !} \;
\ldots  \; \frac{{\mathbf H}_{m, m+1}^{n_m}}{ n_m !}\right\rangle_g
\end{eqnarray}
with $ 2 \leqslant m \leqslant n$.

Equation~(\ref{2.10}) tells us that for Hamiltonians satisfying
condition~(\ref{2.8}), the function $H_{1,m}^{(n)}$ has a
contribution from $H_{1,m-1}^{(n-1)}$, which are
calculated to obtain the coefficient of order $\beta^{n-1}$
in the expansion of the HFE. The number of terms to be
calculated in each order of $\beta$ in this expansion
is then remarkably reduced.

The result~(\ref{2.10}) permits us to write:
\begin{eqnarray}   \label{2.11}
  H_{1,n}^{(n)} = (H_{1,1}^{(1)})^n, \hspace{0.5cm}
           n = 1, 2, 3 \ldots
\end{eqnarray}
and
\begin{eqnarray} \label{2.12}
H_{1,n-1}^{(n)} =  H_{1,1}^{(1)} \times H_{1, n-2}^{(n-1)}
  +   \frac{1}{2!} \left\langle {\mathbf H}_{12}^2  \, {\mathbf H}_{23}
 \ldots {\mathbf H}_{n-1, n} \right\rangle_g \,,
\end{eqnarray}
where the $g$-trace
($\langle\ldots\rangle_g$) in equations (\ref{2.10}) and (\ref{2.12})
means
\begin{eqnarray}  \label{2.6}
\left\langle \overset{m}{\underset{i=1} \prod} \; \;
\frac{{\mathbf H}_{i, i+1}^{n_i}}{n_i !} \right\rangle_g \equiv
 \frac{1}{n!} \;\;  {\underset{\mathcal P} \sum}
   \left\langle {\mathcal P} \left({\mathbf H}_{1,2}^{n_1}, {\mathbf H}_{2,3}^{n_2}, \ldots,
       {\mathbf H}_{m,m+1}^{n_m}\right) \right\rangle  ,
\end{eqnarray}
where $\sum_{i=1}^m n_i = n$
and $n_i \not= 0$ with $i= 1, 2, \ldots, m$. The notation
${\mathcal P} \big({\mathbf H}_{1,2}^{n_1}, {\mathbf H}_{2,3}^{n_2}, \ldots,
{\mathbf H}_{m,m+1}^{n_m}\big)$ means all the distinct permutations
of the $n$ operators where $m$ of them,
$\big\{{\mathbf H}_{1,2}, {\mathbf H}_{2,3}, \ldots,
{\mathbf H}_{m,m+1}\big\}$, are different. The notation
$\langle \ldots \rangle$ corresponds to calculating the
normalized traces on the indexes:
$1, 2, \ldots, m+1$~\cite{chain_m,winder}.

The results~(\ref{2.10})--(\ref{2.12}) are valid for any 1D
Hamiltonian that is invariant under space translation, satisfies the
periodic space condition  and the condition~(\ref{2.8}).

In this article we study a 1D model that satisfies the necessary
conditions to apply the method of reference~\cite{chain_m} plus
the condition~(\ref{2.8}). All the terms in this Hamiltonian
are commutative; hence,  the $g$-traces in~(\ref{2.10})--(\ref{2.12})
can be replaced by the usual normalized traces~\cite{chain_m}.

We calculate the high-temperature expansion of ${\mathcal W}_s (J, h, D; \beta)$,
for arbitrary value of the spin $S$ , up to order $\beta^{17}$.
The relation~(\ref{3.8}) can be  applied to yield the
high-temperature expansion of
${\mathcal W}_S^{\prime} (J^{\prime}, h^{\prime}, D^{\prime}; \beta)$
from the expansion of ${\mathcal W}_s (J, h, D; \beta)$.

Here we present the high-temperature expansion of ${\mathcal W}_s (J, h, D; \beta)$,
for any spin value $S$,  up to order~$\beta^2$,
\begin{eqnarray}  \label{3.9}
{\mathcal W}_s (J, h, D; \beta) & = & -\frac{\ln(2 S+1)}{\beta}
+ \frac{D}{3}
+ \left[- \frac{1}{6} \; \frac{S h^2}{S+1} + \frac{1}{30}  \;  \frac{D^2}{S (S+1)}
- \frac{2}{45}  \; \frac{S D^2}{S+1}
         \right. \nonumber \\
%
&&
 -  \left. \frac{1}{18}  \;  \frac{J^2}{S+1}
- \frac{2}{45}  \;  \frac{D^2}{S+1} - \frac{1}{18}  \; \frac{S J^2}{S+1}
- \frac{1}{6}  \; \frac{h^2}{S+1} \right] \beta
+ \left[ \frac{1}{135}  \; \frac{J^2 D}{(S+1)^2}
      \right.  \nonumber \\
%
&&
 - \frac{1}{30}  \; \frac{h^2 D}{S (S+1)^2} + \frac{2}{45}  \;  \frac{S^2 h^2 D}{(S+1)^2}
+ \frac{4}{45}  \;  \frac{S h^2 D}{(S+1)^2} + \frac{2}{9}  \;  \frac{J S h^2}{(S+1)^2}
- \frac{1}{45}  \;  \frac{J^2 D}{S (S+1)^2}
                \nonumber \\
%
&&
 + \frac{1}{90}  \; \frac{h^2 D}{(S+1)^2}
+ \frac{8}{135}  \; \frac{S J^2 D}{(S+1)^2} - \frac{4}{405}  \; \frac{D^3}{(S+1)^2}
+ \frac{1}{9}  \;  \frac{J h^2}{(S+1)^2}  + \frac{4}{135}  \; \frac{S^2 J^2 D}{(S+1)^2}
                \nonumber  \\
%
&&
 +   \frac{1}{126}  \; \frac{D^3}{S^2 (S+1)^2}
+ \frac{1}{9}  \; \frac{J S^2 h^2}{(S+1)^2}
- \frac{4}{315}  \; \frac{D^3}{S (S+1)^2} + \frac{16}{2835}  \; \frac{S D^3}{(S+1)^2}
                      \nonumber \\
%
&& +  \left. \frac{8}{2835}  \; \frac{S^2 D^3}{(S+1)^2} \right] \beta^2
+ {\cal O} (\beta^3).
\end{eqnarray}

We should note that this high-temperature expansion is valid for positive,
null or  negative  values of $J$, and for arbitrary values
of $S, h$ and $D$. The HFE of the large-spin limit model of Hamiltonian
(\ref{3.2}) is calculated from the high-temperature expansion of
${\mathcal W}_s (J, h, D; \beta)$ by taking the
limit $S \rightarrow \infty$.

The authors maintain a website\footnote{\href{http://www.proac.uff.br/mtt}{http://www.proac.uff.br/mtt}.}
in which the interested reader may find data files
on  the arbitrary finite spin-$S$ and the large-spin limit
($S\rightarrow \infty$)  HFE's
of the normalized Hamiltonian (\ref{3.2}) up to order
$\beta^{17}$.

We have a few general comments on the function
${\mathcal W}_s (J, h, D; \beta)$:

\noindent {\it i}) the expansion (\ref{3.9}) of the HFE is an
even function of the external longitudinal magnetic field $h$;

\noindent {\it ii}) the function ${\mathcal W}_s (J, h, D; \beta)$
is even in the parameter $J$ for an external magnetic field with null
longitudinal component ($h = 0$);

\noindent {\it iii}) even for $h=0$, the HFE (\ref{3.9}) is sensitive
to the sign of the parameter $D$.

By direct comparison, we verify that our expansion
of the HFE of the spin-$S$ Ising model coincides
for $S= 1/2$ and $S=1$ with the high-temperature expansions
of the exact results of references~\cite{bjp01} and~\cite{krinsky}, respectively.

It is simple to understand why the second comment  is valid
for the exact expression of the HFE of the model~(\ref{3.2}).
The partition function  comes from the calculation of
the traces of operators $({\mathbf H}_s)^n$. In reference~\cite{ejp} we showed
that for any spin-$S$ only $\mathrm{tr}_i \left(s_i^z\right)^{2l} \not= 0$.
In the absence of the longitudinal component of the  magnetic field
($h=0$), only products with even number  of operator $J s_i^z s_{i+1}^z$
give non-null contributions to the $\mathrm{tr}[({\mathbf H}_s)^n]$.

The contribution of the single-ion anisotropy term in Hamiltonian
(\ref{3.2}) to the partition function comes from the
operator $\exp\big\{-{\big[\beta D \left(S_i^z\right)^2\big]}\big/{\big[S(S+1)\big]}\big\}$. For positive
values of the crystal field  ($D>0$) the main contribution
of this operator to the partition function comes
from  the smallest eigenvalues of $\left(S_i^z\right)^2$.
On the other hand, for negative values of $D$ the states with
the largest eigenvalues of $\left(S_i^z\right)^2$ are favored. Such
distinct behavior for positive and negative values of $D$
explains the origin of condition ({\it iii}).


\section{The ferromagnetic and anti-ferromagnetic models in the
presence of a weak magnetic field} \label{S4}

The exact expression of the HFE of the Hamiltonian (\ref{3.2}) and
its thermodynamic functions are unknown for an arbitrary value of spin
$s$. Our work has been that of calculating the high-temperature expansion of
thermodynamic functions of one-dimensional models.
In section \ref{S3} we mentioned that the HFE's
of the Hamiltonians (\ref{3.1}) and
(\ref{3.2}) in a   magnetic field with null longitudinal
component  ($h=0$), is an even function of $J$. As a consequence,
for $h=0$, several thermodynamic functions are the
same for the ferromagnetic ($J<0$) and AF ($J>0$)
models, namely: the specific heat, the internal energy, the entropy
and the mean value of the square of the $z$ component of the
spin $\left\langle \left(S_z\right)^2\right\rangle$. For $h=0$, we also have
that the correlation function between first neighbors
satisfies the equality
\begin{eqnarray}   \label{4.1}
{\mathcal C}_s ( - J, 0, D; \beta) =  - \; {\mathcal C}_s (J, 0, D; \beta)
\end{eqnarray}
for $s = \frac{1}{2}, 1, \frac{3}{2}, 2, \ldots$.

The high-temperature expansion (\ref{3.9}) of the HFE  of the
spin-$s$ model is valid for arbitrary values of $h$.
In order to verify how important the presence of a
longitudinal magnetic field is to the thermodynamic
properties of the ferro and AF models,
in this section we examine how the thermodynamic functions of
the ferromagnetic and AF models differ from
their respective values at $h=0$  when they are in the
presence of a weak magnetic field for different values of spin.

Throughout  this section we consider $J= 1$
(AF model) or $J= -1$ (ferromagnetic
model). Again, the parameters are in units of $|J|$ and the
expansions are in powers of $(|J| \beta)$.

Let ${\mathcal F}_s (J, h, D; \beta)$ be a given thermodynamic
function with spin-$s$ (see relation (\ref{3.3})) derived from
Hamiltonian (\ref{3.2}). Its percentage
difference to the value of ${\mathcal F}_s (1, 0, D; \beta)$ is
defined as
\begin{eqnarray}  \label{4.2}
\Delta {\mathcal F}_s (J, h, D; \beta) \equiv
\left[ \frac{{\mathcal F}_s (1, 0, D; \beta) - {\mathcal F}_s (J, h, D; \beta)} {{\mathcal F}_s (1, 0, D; \beta)}  \right] \, \times \, 100\% ,
\qquad J = \pm 1.
\end{eqnarray}

\begin{figure}[!b]
\begin{center}
\includegraphics[width=0.48\textwidth]{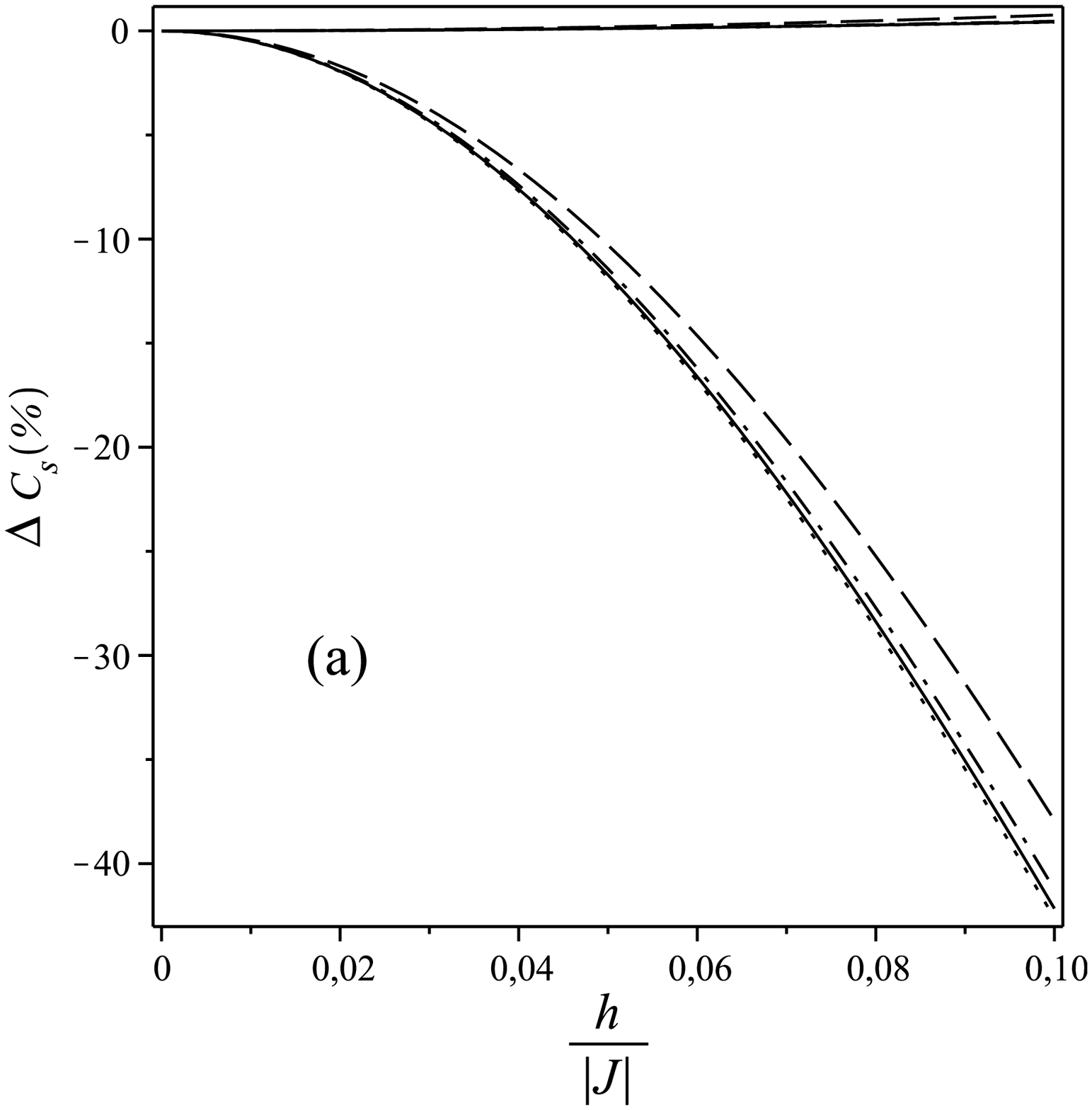}
\hfill
\includegraphics[width=0.48\textwidth]{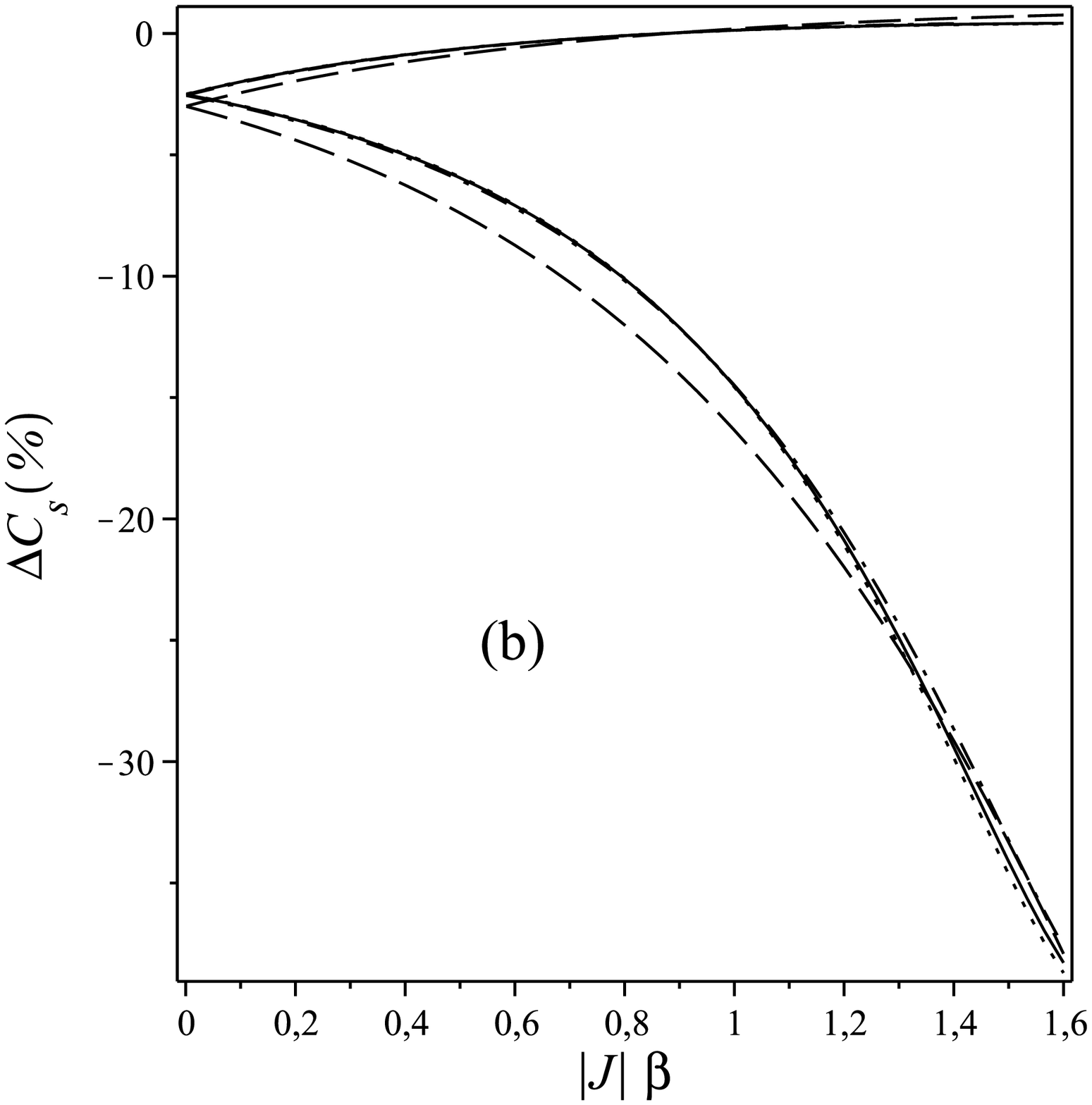}
\vspace{-0.2cm}
\caption{The percentage difference of the specific heat
for $h=0$ and for a weak magnetic field $h$
 for the AF ($J=1$)  and for the ferromagnetic  models ($J= -1$).
Notice that for the AF case, all curves are close
to the horizontal axis.
This comparison is made for both models with $s= 1/2$ (dashed line), $s=2$ (dash-dotted line), $s=4$ (solid line) and the
large-spin limit model (dotted line). In figure~\ref{fig_1}~(a) this percentage difference is
plotted as a function of ${h}/{|J|}$ with  $D = 0$ and
$(|J| \beta) =1.6$. Figure~\ref{fig_1}~(b) presents the curves of $\Delta C_s$ versus
$(|J| \beta)$  with ${D}/{|J|} = - 0.5$ and ${h}/{|J|} = 0.1$.
}
\label{fig_1}
\end{center}
\end{figure}

Let us compare the thermodynamic functions.

\vspace{0.2cm}

1) Comparison of the specific heat per site:
$C_s (J, h, D; \beta)$.

For each spin-$s$, the specific heat functions of the
ferromagnetic and AF models,
in the high temperature limit ($\beta \rightarrow 0$), reach the same
value for any $h$ and $D$. The $\beta^2$ term in this thermodynamic function
has a $J^2$ dependence for both models.

Figure~\ref{fig_1} show $\Delta C_s (\pm 1, {h}/{|J|},
{D}/{|J|}; |J|\beta)$ for $s = 1/2, 2, 4$ and $\infty$
(large-spin limit model) as a function of ${h}/{|J|}$ and
$(|J| \beta)$.  In figure~\ref{fig_1} (a) we have
$D=0$, $|J| \beta =1.6$ and ${h}/{|J|} \in [0, 0.1]$. For these
spin values, \linebreak $|\Delta C_s ( - 1, {h}/{|J|}, 0; 1.6)\lesssim 43\% $ for the ferromagnetic models, whereas
$|\Delta C_s ( 1, {h}/{|J|}, 0; 1.6)| \lesssim 0.42\%
$ for the AF models. In the AF case
the function $\Delta C_s ( 1, {h}/{|J|}, 0; 1.6)$ for $s= 1/2$
very closely approximates its large-spin limit ($s\rightarrow \infty$)
version. Figure~\ref{fig_1} (b) shows the percentage difference $\Delta C_s$ as
a function of $(|J| \beta$), for ${D}/{|J|} = -0.5$ and
${h}/{|J|} = 0.1$. Again  we have that the percentage differences
of the ferromagnetic models are high
($|\Delta C_s ( - 1, 0.1, - 0.5; |J|\beta)| \lesssim 38\% $
whereas the percentage differences for the AF models
are much smaller
($|\Delta C_s ( 1, 0.1, - 0.5; |J|\beta)| \lesssim 3\% $).

The previous discussion  exemplifies  the fact that the
high-temperature expansion  of the specific heat of the AF model
in the presence of a weak longitudinal magnetic field  can be approximated
by the corresponding expression derived from the results of reference~\cite{winder}
(where we have $h=0$). Figure~\ref{fig_1} shows us that for arbitrary spin-$s$
we can improve this  approximation  by recognizing that
$C_s ( 1, {h}/{|J|},{D}/{|J|}; |J|\beta) \approx
C_s ( 1, 0,{D}/{|J|}; |J|\beta) + \Delta C_{1/2} ( 1, {h}/{|J|},{D}/{|J|}; |J|\beta) $, at least in the region in which $ {h}/{|J|} \ll 1$ and
$|J| \beta \lesssim 1.6$.  It is important to recall that the exact expression of
$C_{1/2} (J, h, D; \beta)$ is known~\cite{bjp01}.

For the spin-$s$ ferromagnetic model~(\ref{3.2}) we need to know
the high-temperature expansion  of the HFE in the presence
of longitudinal magnetic field to obtain the value
of $C_s ( -1, {h}/{|J|},{D}/{|J|}; |J|\beta),$ even in the
presence of a weak field.

\vspace{0.2cm}

2) Comparison of the correlation function between
first neighbors per site: ${\mathcal C}_s (J, h, D; \beta)$.

Equation~(\ref{4.1}) tells us that in the absence of magnetic field
($h=0$),
\begin{eqnarray}   \label{4.4}
{\mathcal C}_s ( - J, 0, D; \beta) =  - \; {\mathcal C}_s (J, 0, D; \beta)
\end{eqnarray}
for $s = \frac{1}{2}, 1, \frac{3}{2}, 2, \ldots$
Result~(\ref{4.4}) describes the parallel
(anti-parallel) alignment of the neighboring spins
in the ferromagnetic (AF) model.

We want to check how the function
${\mathcal C}_s ( \pm 1, {h}/{|J|}, {D}/{|J|}; |J|\beta)$
differs from the equality  (\ref{4.4}) in the presence of a weak
magnetic field. In order to quantify this deviation, we define
\begin{eqnarray}  \label{4.5}
{\mathcal D} {\mathcal C}_s \left({h}/{|J|}, {D}/{|J|}; |J|\beta\right) \equiv
\left[\frac{{\mathcal C}_s \left( - 1, {h}/{|J|}, {D}/{|J|}; |J| \beta\right)
+ {\mathcal C}_s \left( 1, {h}/{|J|}, {D}/{|J|}; |J| \beta\right)}
     {{\mathcal C}_s \left( - 1, {h}/{|J|}, {D}/{|J|}; |J| \beta\right)}\right]
 \, \times \, 100\% \,.
\end{eqnarray}

\begin{figure}[!b]
\begin{center}
\includegraphics[width= 5.5cm,height= 6.5cm,angle= 0]{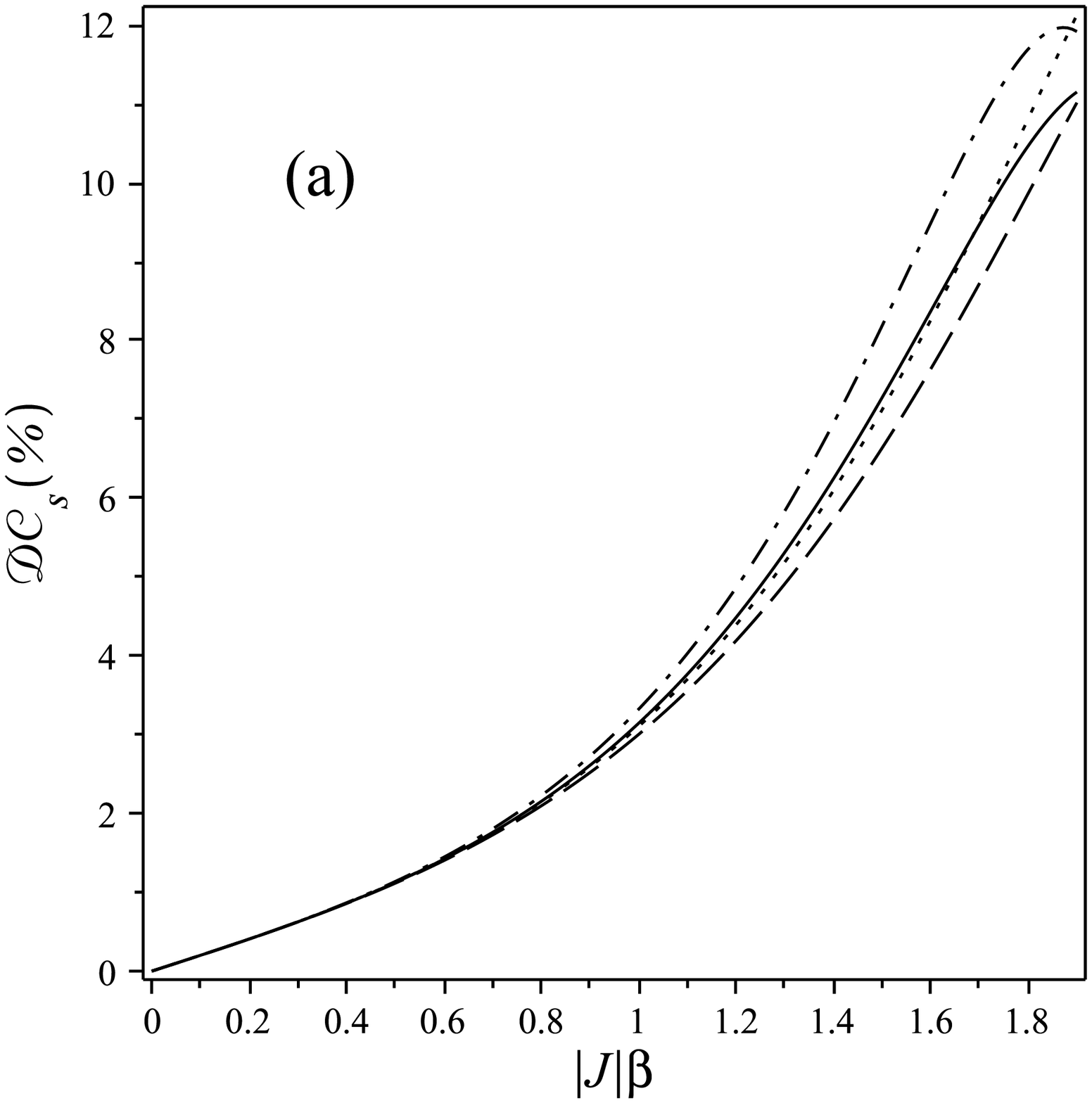}
\hspace{1cm}
\includegraphics[width=5.5cm,height= 6.5cm,angle= 0]{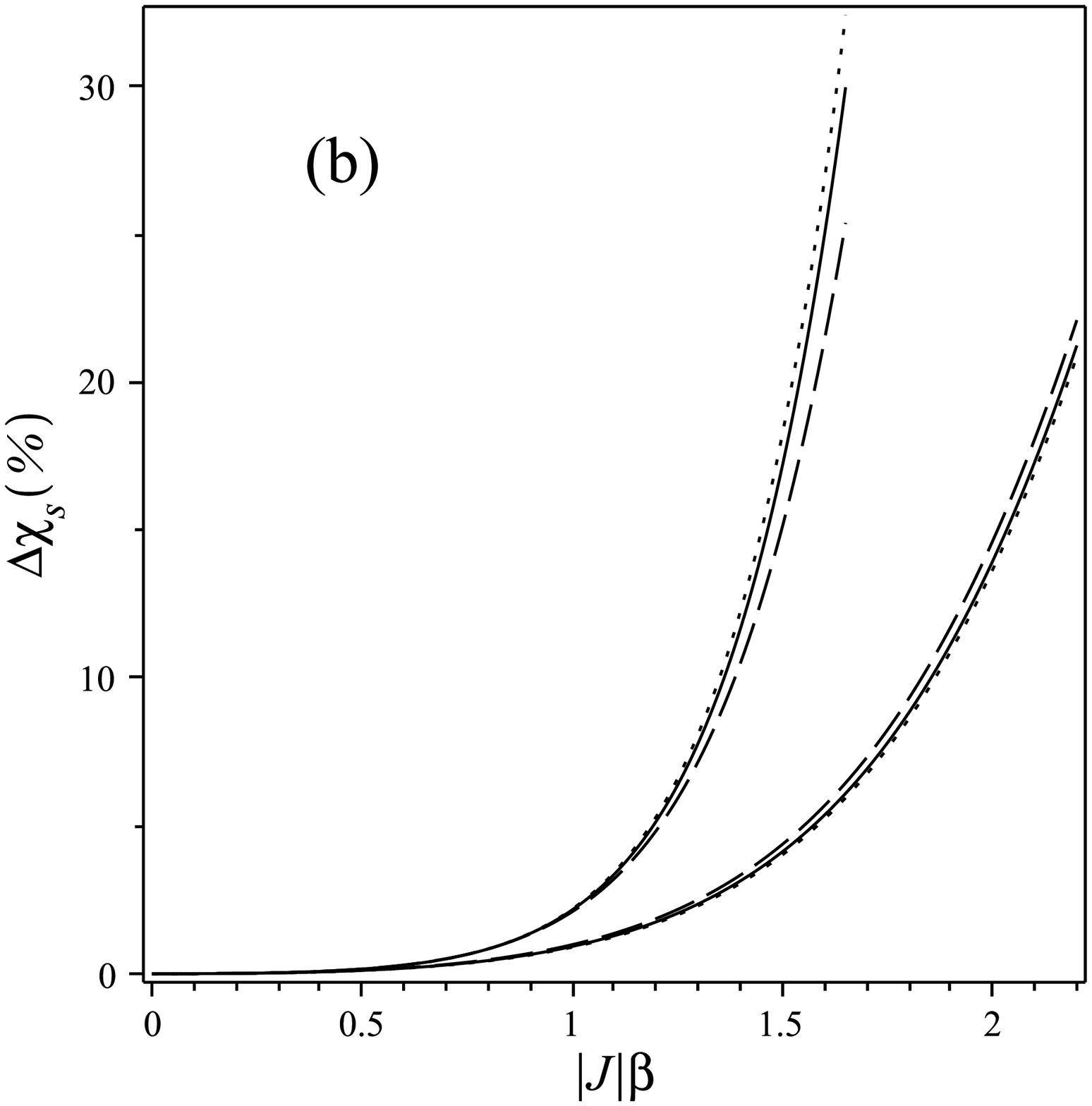}
\vspace{-0.2cm}
\caption{
Left: comparison of the percentage function ${\mathcal D} {\mathcal C}_s$
for the $s=2$ and the large-spin limit ($s \rightarrow \infty$)
models for ${h}/{|J|} = 0.1$: for $s=2$, the cases $D=0$ (dashed line)
and ${D}/{|J|} = - 0.5$ (solid line) are shown, whereas for the
$s \rightarrow \infty$ model the $D=0$ (dotted line)
and ${D}/{|J|} = - 0.5$ (dash-dotted line) cases are shown.
Right: comparison of the percentage
difference of the magnetic susceptibility $\Delta \chi_s (-1, 0.1, D; \beta)$
in the ferromagnetic  model ($J = -1$) for
$s\in\{2,4,\infty\}$,
for ${D}/{|J|} = - 0.5$ (upper set of curves)
and for ${D}/{|J|} = - 0.5$ (lower set of curves);
in both sets, the $s=2$ (dashed line), $s=4$ (solid line) and $s\rightarrow \infty$ (dotted line)
cases are shown.}
\label{fig_2}
\end{center}
\end{figure}

In figure~\ref{fig_2} (a) we show the percentage function~(\ref{4.5}),
with $D=0$ and ${D}/{|J|} = - 0.5$, with ${h}/{|J|} = 0.1$,
to verify the departure from equation~(\ref{4.4}) for the
ferromagnetic and AF models in the presence of
a weak magnetic field.  The function
${\mathcal D} {\mathcal C}_s ( {h}/{|J|}, {D}/{|J|}; |J|\beta)$ depends on the
spin value and for lower temperatures ($|J| \beta \sim 1.9$) it
can be around $~10\%$.

\vspace{0.2cm}

3) Comparison of the magnetic susceptibility per site:
$\chi_s (J, h, D; \beta)$.

It is simple to obtain the high-temperature expansion of the magnetic
susceptibility per site, \linebreak $\chi_s (J, h, D; \beta)$, from the
series~(\ref{3.9}) for the HFE
$\left(\chi_s (\beta) =  - {\partial^2 {\mathcal W}_s (\beta)}/{\partial h^2}\right)$.

At $h=0$, the magnetic susceptibility of the ferromagnetic
and AF  models are distinct. We define a percentage
difference analogous to equation~(\ref{4.2}) to
present in figure~\ref{fig_2}~(b) the difference between
$\chi_s (-1, 0.1, {D}/{|J|}; |J| \beta)$ and
$\chi_s (-1, 0, {D}/{|J|}; |J| \beta)$ in
the ferromagnetic models with $s = 2, 4$ and the large-spin limit model. In
this figure we have ${D}/{|J|} = \pm 0.5$. For lower temperatures
($|J| \beta \sim 2)$ we have percentage differences around $20\%$.
By looking at figure~\ref{fig_2}~(b) we see that for
$J= -1$ we have
${\rd(\Delta \chi_s)}/{\rd\beta}\big|_{D= -0.5}
> {\rd(\Delta \chi_s)}/{\rd\beta}\big|_{D= 0.5}$\,. This
inequality  is explained by the fact that for $D<0$ the states
with the largest  values of $\left(S_i^z\right)^2$ are more probable,
whereas the states with the smallest values of
$\left(S_i^z\right)^2$ are favored in the model with $D>0$.

For the AF models ($J=1$), in the interval
$|J| \beta \in [0, 1.43]$, with ${h}/{|J|} = 0.1$
and ${D}/{|J|} = \pm 0.5$, we obtain from the percentage
difference~(\ref{4.2})
$|\Delta \chi_s (1, 0.1, {D}/{|J|}; |J| \beta)| \lesssim 0.6\%$.

We point out that to derive the high-temperature expansion of
$\chi_s (J, h, D; \beta)$ from the HFE, we need the
dependence of the HFE on $h$. That is not the case
of the previous paper on the spin-$S$ Ising model~\cite{phys07}.


\section{Thermodynamic behavior of the single-chain  magnet in
the interval $10~\text{K} \lesssim T \lesssim 40~\text{K}$} \label{S5}

The Hamiltonian~(\ref{3.1}), with $S=3$, was applied by
Kishine  et al.~\cite{kishine} to analyze the low energy
dynamics of the single-chain magnet (SCM)
$[{\rm Mn (saltmen)}]_2[{\rm Ni (pao)}_2{\rm (py)}_2] {\rm (PF}_6)_2$
for $T< 40$~K in the presence of a weak magnetic field.
Their Hamiltonian~(2) is identical to ours~(\ref{3.1})
once we set $J^{\prime} = - 2t$.

In~\cite{ejp} we studied the thermodynamics of the
unitary spin-$s$ $XXZ$ model  with a single-ion anisotropy term in
the presence  of a magnetic field in the $z$ direction. We obtained
that the specific heat, the magnetization and the magnetic
susceptibility of this model with $s=3$ are well approximated
by their respective large-spin limit versions, in the temperature
range of $|J|\beta \lesssim 1$.
For the non-normalized $S=3$ $XXZ$ model this range
corresponds to $|J|\beta \lesssim 0.083$
(see section~2.2 of reference~\cite{ejp}).

In this section, we compare the Hamiltonian~(\ref{3.1})
having spin-$3$ with its large-spin limit version
for $T \lesssim 40$~K.
By ``large-spin limit model'' we mean
that the z component of its spin vector
$\vec{S}$ varies continuously, namely, $S_i^z = 2\sqrt{3}
\cos(\theta_i)$, in which $\theta_i \in [0, \pi]$,
and $i \in \{ 1, 2, \ldots, N\}$. In order to relate our
analysis to the aforementioned SCM we
use the same parameter values as in~\cite{kishine},  namely,
\begin{subequations}
\begin{eqnarray}  \label{5.1a}
 \frac{J^{\prime}}{k} = -1.6~\text{K}
\end{eqnarray}
and
\begin{eqnarray}  \label{5.1b}
 \frac{D^{\prime}}{k} = -2.5~\text{K} ,
\end{eqnarray}
\end{subequations}
in the spin-3 and large-spin limit Hamiltonians.
One is reminded that $k$ is the Boltzmann constant.
Taking the value~(\ref{5.1a}) for $J^{\prime}$,
for $T\sim 40$~K we have $|J^{\prime}| \beta \sim 0.04$.
Note that the temperature region characterized
by $|J|\beta \lesssim 0.04$
is contained in the temperature region in which the spin-3 $XXZ$
model behaves very much like its large-spin limit version.

Along this section, we take ${h^{\prime}}/{k} = 0.25$~K,
which is a weak magnetic field (${h^{\prime}}/{|D^\prime|} = 0.1$).

Let ${\mathcal F}_S (h; \beta)$ be a thermodynamic function.
Its  percentage difference of the $S=3$ and the
large-spin limit models is defined as
\begin{eqnarray}  \label{5.2}
\Delta {\mathcal F} (h; \beta) \equiv
\left[\frac{{\mathcal F}_{S\rightarrow \infty} (h; \beta)
    - {\mathcal F}_3 (h; \beta)} {{\mathcal F}_{S\rightarrow \infty} (h; \beta)}  \right]
\; \times \; 100\% \, .
\end{eqnarray}

\begin{figure}[!ht]
\begin{center}
\includegraphics[width= 5.5cm,height= 6.5cm,angle= 0]{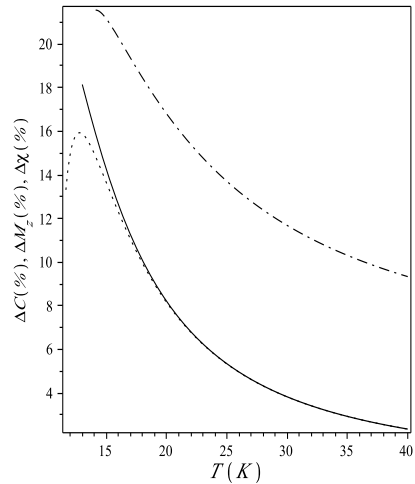}
\vspace{-0.2cm}
\caption{ The percentage differences between the thermodynamic functions
of the spin-$3$ model and its large-spin limit version of the SCM. The
dash-dotted line corresponds to $\Delta C (0.25; \beta)$,  the
solid line to $\Delta M_z (0.25; \beta)$ and the
dotted line to $\Delta \chi (0.25; \beta)$. }
	    \label{fig_3}
\end{center}
\end{figure}

In figure~\ref{fig_3} we plot the percentage difference~(\ref{5.2}) of the
specific heat, the $z$ component of the magnetization per site
$M_z^{(S)} (J^{\prime}, h^{\prime}, D^{\prime}; \beta)$
$\left( M_z^{(S)} (\beta) \equiv - \frac{\partial {\mathcal W}_S (\beta)}{\partial h} \right)$
and the magnetic susceptibility per site.
The  dash-dotted curve corresponds to the specific
heat in the interval $T \in [14.1~\text{K}, 40~\text{K}]$. For
$T  \lesssim 40$~K
or $|J^{\prime}| \beta \gtrsim 0.046$,
we verify that
$\Delta C (0.25; \beta) \gtrsim 10\%$. The percentage
difference of the $z$ component of the magnetization,
$\Delta M_z (0.25; \beta)$, is given by
the solid line in the figure for $T \in [11.63~\text{K}, 40~\text{K}]$, while the
dotted curve describes $\Delta \chi (0.25; \beta)$ in the
temperature window $T \in [12.99~\text{K}, 40~\text{K}]$. We verify that for
$T \lesssim 19.3$~K,
that is, $|J^{\prime}| \beta \gtrsim 0.083$,
 the $z$ component of the
magnetization and the magnetic susceptibility of the
spin-$3$ model of the SCM differ more than $8.5\%$ from their
corresponding large-spin limit values.

Due to the ratio  ${|J^{\prime}|}/{|D^\prime|}$
value of $0.64$ for this SCM, the single-ion anisotropy term
gives the main contribution to its thermodynamics.
Since the crystal field $D^{\prime}$  is
negative (see equation~(\ref{5.1b})), the states with
$S_i^z = \pm 3$ and $\pm2$ are the most probable. The modulus
of the amount of energy, in units of $k$, for the single-ion
anisotropy  term  to change from states with $S_i^z = \pm 2$ to
states with  $S_i^z = \pm 3$, and vice-versa, is $12.5$~K. This
value is close to one-third of 40~K. We conclude that for
$T \lesssim 40$~K
(that is, $|J^{\prime}|\beta \gtrsim 0.04$),
the discretized nature of the spin-$3$ is still an important
feature of the model. This result is very different from that of
reference~\cite{ejp} for the spin-$3$ $XXZ$ model  with a
single-ion anisotropy term in the presence  of a magnetic field
in the $z$ direction.

Our expansion of the HFE with $S=3$ can be easily
used to fit the experimental data of this SCM. It can also be
applied to determine the best fit of the parameters
$J^{\prime}$  and $D^{\prime}$.


\section{Conclusions} \label{S6}

The method of reference~\cite{chain_m} permits one to calculate the
high-temperature expansion of the HFE of any chain Hamiltonian with
interaction between first neighbors which is invariant under space
translation and satisfies a periodic space condition. If the
Hamiltonian also satisfies the condition~(\ref{2.8}), we show that
some of the terms that contribute to  the function
$H_{1,m}^{(n)}$  have already been
calculated in  a lower order of $n$, that is,
$H_{1,m-1}^{(n-1)}$. A set of important 1D Hamiltonians
satisfy the condition~(\ref{2.8}). This contribution from lower
order of $n$ to this auxiliary function  permits us
to compute the HFE of the one-dimension Ising model
with single-ion anisotropy term in the
presence of a longitudinal magnetic field  up to order
$\beta^{17}$, for arbitrary values of spin $S$ and of the other
parameters $J$, $h$ and $D$.  Upon performing
the numerical analysis of the thermodynamics of
spin models~(\ref{3.1}) and~(\ref{3.2})  one must know
the value of the spin beforehand.

We discuss the thermodynamics  of the normalized
Hamiltonian (see equation~(\ref{3.2})), but equation~(\ref{3.8})
relates the HFE of the Hamiltonian~(\ref{3.1}) and its
normalized version (equation~(\ref{3.2})).

Some thermodynamic functions are insensitive
to the sign of $J$ in the absence of a magnetic
field; for instance, the specific heat of the ferromagnetic
($J<0$)  and  AF ($J>0$) spin-$S$ Ising
models with single-ion anisotropy term are identical
with $h=0$. In the absence of a magnetic field,
the ferromagnetic model favors parallel
neighboring spins while in the AF
model the anti-parallel pairs are more probable.
The effect of the presence of an external
magnetic field in both models  is that of favoring the alignment of spins at
each site to the field direction.  In the
ferromagnetic model such alignment is favored
by the coupling between neighboring spins. On the other
hand, in the AF  model there is a
competition between the coupling of neighboring spins
(favoring anti-parallel alignment) and the Zeeman term
(forcing all the spins in the chain to align
with the external magnetic field). As a consequence
of this competition,  there is a  perturbative effect
in the AF model due to the presence of a
weak magnetic field; in such regime,
their thermodynamic functions
${\mathcal F}_s (1, {h}/{|J|}, {h}/{|J|}; |J| \beta)$
for the spin-$s$ model can be approximated by
${\mathcal F}_s (1, {h}/{|J|}, {D}/{|J|}; |J| \beta)
\approx {\mathcal F}_s (1, 0, {D}/{|J|}; |J| \beta)
+ \Delta {\mathcal F}_{1/2} (1, {h}/{|J|}, {h}/{|J|}; |J| \beta)$.
The exact expression of the  HFE of the spin-$1/2$ Ising
model with a single-ion anisotropy term in the presence of a
longitudinal magnetic field is known~\cite{bjp01}. In the
ferromagnetic model, the effect of the Zeeman
coupling cannot be treated anymore as a perturbation
to the interaction between first neighbors and to the
single-ion anisotropy term for
${h}/{|J|} \gtrsim 0.04$.

Kishine  et al.~\cite{kishine} applied the spin-$3$
Hamiltonian~(\ref{3.1}) to analyse  the low energy dynamics
of the \linebreak ${\rm [Mn (saltmen)]_2 [Ni(pac)_2 (py)_2] (PF_6)_2}$
SCM  for temperatures $T<40$~K.
Although $S=3$ could be considered a high spin value
in the region of $T\sim 40$~K ($|J^{\prime}|\beta \sim 0.04$),
the negative value of $D^{\prime}$ favors the states with
$S_i^z = \pm 2$ and $\pm 3$. The modulus of the amount
of energy required by the single-ion anisotropy term
to have $S_z$ varied from $\pm 2 \rightleftharpoons \pm3 $
and vice-versa is about one-third of the
thermal energy available for $T \lesssim 40$~K.
Our results show that the large-spin limit
model, with the spin-3 replaced by the classical vector in the
Hamiltonian, yields a poor approximation to the behavior of this
SCM for $T \lesssim 18$~K and for the set of
parameters values~(\ref{5.1a}) and~(\ref{5.1b}).

Finally, it is very important to point out
that our high-temperature expansion of the HFE  of the spin-$s$
Ising model can be applied  to fit experimental data
of new materials with one-dimensional behavior and
strong anisotropy axis. This expansion leaves the spin value
of the material  as one of the parameters to be determined
by the best fit. The expansion is valid for positive
and negative values of $J$ and $D$  in the presence of a
longitudinal magnetic field that does not have to be a
weak field.

\section*{Acknowledgements}

M.T. Thomaz   thanks  CNPq (Fellowship CNPq, Brazil,
Proc. No.:~30.0549/83--FA) and FAPEMIG for the partial financial support.
O.R. thanks FAPEMIG and CNPq  for the partial financial  support.
The authors are in debt with E.V. Corr\^ea Silva for the careful
reading of the manuscript.


\ukrainianpart

\title{Високотемпературний розклад для класичної моделі Ізінга із членом  $S_z^2$ }

\author{М.Т. Томас\refaddr{ad1}, О. Рохас\refaddr{ad2}}

\addresses{\addr{ad1} Інститут фізики, Федеральний університет Флуміненсе, Нітерой-RJ, Бразилія
\addr{ad2} Факультет точних наук, Федеральний університет м.~Лаврас,  Лаврас-MG,  Бразилія
}

\makeukrtitle

\begin{abstract}
\tolerance=3000%
Ми виводимо високотемпературний розклад вільної енергії Гельмгольца до членів порядку
$\beta^{17}$  для одновимірної $S$-спінової моделі Ізінга, із одноіонною анізотропією в присутності
поздовжнього магнетного поля. Ми показуємо, що значення термодинамічних функцій  феромагнетних моделей
в присутності слабого магнетного поля  не є малими поправками   при  $h=0$.  Ця модель з $S=3$
була застосована Кашіне та ін. [J.-i. Kishine et al.,
Phys. Rev.~B, 2006, \textbf{74}, 224419] для аналізу експериментальних даних одноланцюжкового магнета
${\rm [Mn (saltmen)]_2 [Ni(pac)_2 (py)_2] (PF_6)_2}$ при $T<40$~K.
Ми показуємо, що при $T<35$~K термодинамічні функції моделі в границі великого спіну є поганим наближенням для аналогічних до них $3$-спінових функцій.

\keywords квантова статистична механіка, одновимірна модель Ізінга,  $S$-спінові моделі,
границя великого спіну,  одноланцюжкові магнети
\end{abstract}

\end{document}